\documentstyle[11pt,aaspp4]{article}

\slugcomment{POPe-752, UTAP-285}

\lefthead{Suginohara, Suginohara, \& Spergel}
\righthead{Detecting $z>10$ objects through C, N and O emission lines}

\begin{document}
\title{Detecting $z>10$ objects \\
through carbon, nitrogen and oxygen emission lines}
\author{Maki Suginohara\altaffilmark{1,2},
Tatsushi Suginohara\altaffilmark{1,3,4},
and David N. Spergel\altaffilmark{5}}
\affil{Department of Astrophysical Sciences, Princeton University,
Princeton, NJ 08544}

\altaffiltext{1}{Department of Physics, the University of Tokyo,
Tokyo 113, Japan}
\altaffiltext{2}{makis@astro.Princeton.EDU, JSPS Research Fellow}
\altaffiltext{3}{RESCEU, School of Sciences, the University of Tokyo,
Tokyo 113, Japan}
\altaffiltext{4}{tatsushi@astro.Princeton.EDU, JSPS Postdoctoral Fellow}
\altaffiltext{5}{dns@astro.Princeton.EDU}
\begin{abstract}
By redshift of $10$, star formation in the first objects
should have produced
considerable amounts of Carbon, Nitrogen and Oxygen.
The submillimeter lines of C, N and O redshift into
the millimeter and centimeter bands ($0.5\,{\rm mm}$--$1.2\,{\rm cm}$),
where they may be detectable.
High spectral resolution observations could potentially detect
inhomogeneities in C, N and O emission,
and see the first objects forming at high redshift.
We calculate expected intensity fluctuations and discuss
frequency and angular resolution required to detect them.
For CII emission, we estimate the intensity using two independent
methods: the line emission coefficient argument and
the luminosity density argument.
We find they are in good agreement.
At $1+z \sim 10$, the typical protogalaxy has a velocity dispersion of 
$30 \,{\rm km\,s^{-1}}$ and angular size of $1 \,{\rm arcsecond}$.
If CII is the dominant coolant, then we estimate a characteristic line 
strength of $\sim 0.1 \,{\rm K\, km\, s}^{-1}$.
We also discuss other atomic lines and estimate their signal.
Observations with angular resolution of $10^{-3}$
can detect moderately nonlinear fluctuations of amplitude
$2 \cdot 10^{-5}$ times the microwave background.
If the intensity fluctuations are detected,
they will probe matter density inhomogeneity,
chemical evolution and ionization history at high redshifts.
\end{abstract}

\begin{center}
submitted to {\it The Astrophysical Journal}
\end{center}

\clearpage
\section{Introduction}
When did the first stars and galaxies form?
How can we detect them?
Quasars were once the most distant objects that have ever been
observed (\cite{sch91}).
Now galaxies are beginning to take their place.
Discovery of a large number of galaxies at $z>3$ (\cite{ste96}), 
and even up to $z\sim 5$ (\cite{fra97,dey98}), suggests that
galaxies existed well before quasars did.
In hierarchical models, star formation starts relatively
early, by $z\sim30$ (\cite{cou86,fuk94,gne96,teg97}),
so even these $z \sim 5$ galaxies are not representative of the first
generation of objects.

The purpose of this paper is to investigate the possibility 
of detecting primordial objects through the atomic lines
of carbon, nitrogen and oxygen,
which should have been produced as a result of star formation.
At low redshift,
these submillimeter lines are important coolants
(e.~g., \cite{ben94,isr95,stu97,you97,barv97,mal97}).
If the source is at redshift $1+z \sim 10$--$20$,
the lines are redshifted to radio frequencies,
where
the atmosphere is more transparent
and ground--based
detectors with
very high sensitivity and resolution
exist.

If the source was uniformly distributed through space,
then its emission would merely distort the spectrum.
Since heavy elements produced by stars are distributed
inhomogeneously, the observed emission should have
intensity fluctuations
as a function of both position and frequency.
With high enough resolution one can in principle distinguish them
from background.
If we take $1 h^{-1}\,{\rm Mpc}$ as a characteristic comoving
scale of the inhomogeneity,
this corresponds to 
$\sim 1$ arcmin in angular scale,
and $\sim 10^{-3}$ in frequency resolution.
On smaller scales,
velocity dispersion of matter limits the scale that can be
resolved in frequency space.
In this paper we formulate the intensity fluctuations 
and estimate their magnitude.
This signal should give us information on number density
fluctuations of specific source elements at $1+z>10$.
Detailed determination of the peak of this signal will also tell us at 
which epoch elements were ionized.

The rest of this paper is organized as follows.
In the next section,
we present our formalism for estimating the fluctuating signal.
Then we estimate its magnitude in the case that the matter density
fluctuations are calculated from the linear power spectrum
in section \ref{sect:linear}.
We also estimate optimal
frequency and angular resolution
for an object search.
We focus on atomic line redshifted into present radio band,
investigating carbon,
nitrogen and oxygen emission lines at rest frame wavelength of hundreds of
micrometers.
In section \ref{sect:nonlin}, we discuss the size and the number
density of nonlinear clumps at the reionization epoch, and
estimate intensity from nonlinear clumps and
the interval between enhance of intensity
fluctuations.
Section \ref{sect:paremis} gives discussion of
merits and demerits as the signal of specific emission.
After alternative estimation of CII luminosity
in section \ref{sect:altcii},
the final section gives our conclusion.
\section{Observable intensity fluctuations}
\subsection{Formulation}
\label{subsect:formul}
Throughout this paper we make the following assumptions:
by $1+z\sim 10$--$20$, star formation has both formed metals and heated the 
surrounding gas initially to $T\sim 10^4\,{\rm K}$.  
Cooling is compensated by heating so that the gas temperature $T_{\rm gas}$
is maintained much higher than the CMB temperature at that epoch
but lower than $\sim 10^4\,{\rm K}$.
Cooling by atomic transitions exceeds Compton cooling, as we will
see below.

Intensity $I_{\nu}(z)$ of radiation in emitting 
medium at redshift $z$ is obtained by solving the radiative transfer
equation:
\begin{equation}
\label{eqn:radtrans}
   \frac{d I_{\nu}(z)}{d t} = 
   c j_{\nu}(z) - 3 H(z) I_{\nu}(z),
\end{equation}
where $c$ is the speed of light, and $j_{\nu}(z)$ and $H(z)$ denote the
spontaneous emission coefficient and
Hubble parameter, respectively.
Differentiation with respect to time in the left hand side is taken
with $\nu(1+z)^{-1}={\rm const.}$
The second term in the right hand side
comes from the effect of cosmic expansion.
We have assumed that absorption is negligible.
Integrating equation (\ref{eqn:radtrans}) yields the present
intensity at frequency $\nu_0$:
\begin{equation}
\label{eqn:radtra2}
   I_{\nu_0}(z=0) = \int \frac{c j_{\nu}(z)}{(1+z)^3} dt
   = \int \frac{c}{H(z) (1+z)^4} j_{\nu}(z) dz ,
\end{equation}
where $\nu = \nu_0 (1+z)$ in the integrand. 
For spontaneous transition from an upper level $u$ to a lower level $l$
releasing a photon with energy $h \nu_{\rm line}$,
the emission coefficient is 
\begin{equation}
\label{eqn:emcoeff}
   j_{\nu} = \frac{h_{\rm P} \nu_{\rm line}}{4 \pi} n_u
   A_{ul} \phi (\nu),
\end{equation}
where $h_{\rm P}$ is the Planck constant,
$n_u$ is the number density of the source in the upper level,
$A_{ul}$ is the Einstein $A$--coefficient, and
$\phi(\nu)=\delta_D(\nu - \nu_{\rm line})$ is the line profile
function, which we take to be equal to delta function because both
natural and thermal broadening is negligible for the expected frequency
resolution.
Since redshift of the source $z_{\rm s}$ is inferred from the relation:
\begin{equation}
   1+z_{\rm s} = \frac{\nu_{\rm line}}{\nu_0},
\end{equation}
we can express the line profile function as
\begin{equation}
\label{eqn:lineprof}
   \phi (\nu) = \phi \left( \nu_0(1+z) \right)
   = \frac{1+z_{\rm s}}{\nu_{\rm line}}
   \delta_D(z-z_{\rm s}).
\end{equation}
Using equations (\ref{eqn:emcoeff}) and (\ref{eqn:lineprof})
in equation (\ref{eqn:radtra2}) yields intensity due to the line
emission:
\begin{equation}
\label{eqn:intenhub}
   I_{\rm line}
   = \frac{h_{\rm P} c}{4 \pi} A_{ul} n_u H^{-1}(z_{\rm s})
   (1+z_{\rm s})^{-3}.
\end{equation}
In the Einstein--de Sitter universe, equation
(\ref{eqn:intenhub}) reads
\begin{eqnarray}
\label{eqn:inteneds}
   I_{\rm line}
   &=& \frac{h_{\rm P} c}{4 \pi} H_0^{-1}
   A_{ul} f_u n_{\rm a} (1+z_{\rm s})^{-9/2},
\end{eqnarray}
where
$f_u = n_u / n_{\rm a}$ is the fraction of the source that is in the
upper state,
and $n_{\rm a}$ is the local number density of the source.
\subsection{Magnetic dipole transition}
\label{subsect:trans}
The spin--orbit interaction splits each spectroscopic term of an atom
into a set of energy levels.
Magnetic dipole transitions near the ground state of neutral and
ionized carbon, nitrogen and oxygen have wavelength of hundreds of
micrometers.
We list such transitions in table \ref{tab:lines} (\cite{mel88}).
These are candidates which produce intensity fluctuations.

\placetable{tab:lines}
\subsection{Critical number density}
\label{subsect:crnum}
We have seen the line intensity is proportional to
the number density of the source that is in the upper state.
Let us examine the condition for substantial number of the source
to be in the upper state.
If either spontaneous or stimulated emission
occurs more rapidly than
the collisional excitation by surrounding gas (mainly Hydrogen),
very few atoms (or ions) will be in the upper state.
The condition for collisional rate exceeding spontaneous
and stimulated
emission rates is
\begin{equation}
\label{eqn:condit}
   n_{\rm gas} \geq n_{\rm gas}^{\rm cr},
\end{equation}
where $n_{\rm gas}^{\rm cr}$ is the critical number density determined by
\begin{equation}
\gamma_{ul}n_{\rm gas}^{\rm cr}=A_{ul}
\biggl[1-\exp\biggl(
-\frac{h_{\rm P} \nu_{\rm line}}{kT_{\rm CMB}(z_s)}
\biggr)\biggr]^{-1},
\end{equation}
with $\gamma_{ul}$ being the rate coefficient for
collisional de--excitation,
and $T_{\rm CMB}(z)$ the CMB temperature at redshift $z$.

If $n_{\rm gas} > n_{\rm gas}^{\rm cr}$,
the excitation temperature of the source
$T_{\rm source}$ is almost equal to $T_{\rm gas}$,
where $T_{\rm gas}$ is maintained around $10^3\,{\rm K}$.
In this case, the fraction $f_u=f_u^{\rm eq}$
does not depend on $T_{\rm source}$.
In CI case,
the excitation energies of ${}^3P_2$ and ${}^1D$ levels
correspond to $60 {\rm K}$ and $1.5\times10^4 {\rm K}$, respectively.
If we assume that the gas is in thermal equilibrium at a
temperature around $10^3 {\rm K}$,
electrons in the $(2p)^2$ shell of CI easily transit within
${}^3P$ levels, while it is almost impossible to jump to
${}^1D$ level.
Therefore the fraction of atom that is in the state labeled as $u$
is independent of matter temperature and determined by
\begin{equation}
   f_u^{\rm eq} \equiv \frac{n_u}{n_{\rm tot}}
   = \frac{g_u}{\sum_{\rm same \,term} g_i},
\end{equation}
where $g_u = 2J+1$ is the statistical weight of level $u$.
Specifically, the fraction is
$1/9$, $3/9$ and $5/9$ for
${}^3P_0$,${}^3P_1$ and ${}^3P_2$ of neutral carbon.
If $n_{\rm gas} < n_{\rm gas}^{\rm cr}$,
the line intensity is diminished to
\begin{equation}
   I_{\rm line} \simeq
   I_{\rm line}^{\rm eq}
   \left( \frac{n}{n + n^{\rm cr}} \right)_{\rm gas}.
\end{equation}
In CI ($609 \,\mu {\rm m}$) case,
the critical number density of Hydrogen is given by
\begin{equation}
n_{\rm H}^{\rm cr} 
= 73 \biggl(\frac{T_{\rm gas}}{10^3 \, {\rm K}}\biggr)^{-0.34}
\biggl[ 1 - \exp \biggl(-\frac{8.7}{1+z_s}\biggr) \biggr]^{-1}
\, {\rm cm^{-3}}
\end{equation}
(\cite{hol89}).
The last factor in the above expression comes from the effect of
stimulated emission.
The effect is important for the CI $609 \,\mu {\rm m}$ line,
but is negligible for the other fine structure transitions of interest
here at $1+z_s \lesssim 20$.
The excitation of ions is induced mainly by the collisions with
electrons, if the gas is ionized.
The condition (\ref{eqn:condit}) is determined by
the critical number density of electron
\begin{equation}
n_{\rm e}^{\rm cr} 
= 28 \biggl(\frac{T_{\rm gas}}{10^3 \, {\rm K}}\biggr)^{0.50}\, {\rm cm^{-3}}
\end{equation}
(\cite{hol89}),
in CII ($158 \mu {\rm m}$) case.
\subsection{Gas clumping}
At $1+z\sim 10$--$20$, the mean density of gas
$\simeq n_{\rm H}^{\rm mean}=2 \times 10^{-4} [(1+z)/10]^3
\,{\rm cm^{-3}}$
is lower than the critical number density.
The condition for sufficient magnitude of signal, however,
is attached to local number density of the source.
The condition (\ref{eqn:condit}) can hold
in the virialized object,
even if $n_{\rm gas}^{\rm mean} < n_{\rm gas}^{\rm cr}$.

Let us express by $F^{\rm cr}$
the fraction of gas within the relevant scale
which satisfies
$n_{\rm gas} > n_{\rm gas}^{\rm cr}$.
Then the smoothed intensity is given by
\begin{equation}
\label{eqn:smoint}
   \bar{I}_{\rm line}
   = \frac{h_{\rm P} c}{4 \pi} H_0^{-1}
   A_{ul} f_u^{\rm eq} F^{\rm cr}
  \left[ \frac{n_{\rm a}}{n_{\rm H}} (z_{\rm s}) \right]
   n_{{\rm H},0} (1+z_{\rm s})^{-3/2},
\end{equation}
where
$\left[ n_{\rm a}/n_{\rm H} \right]$ is relative abundance of the
atom, and $n_{{\rm H},0}$ is present Hydrogen number density.
It is uncertain to what degree the baryonic gas condenses
in dark matter halos.
However, it seems reasonable to expect that 
the gas inside such halos can radiatively cool and form a gaseous
disk,
whose radius is $\lambda r_{\rm vir}$,
where $\lambda$ is the spin parameter and
$r_{\rm vir}$ is the virial radius (\cite{pee93,dal97}).
If this is the case,
we expect the mean gas density to be at least
$18\pi^2 \lambda^{-3}$ ($\sim 10^6$)
times the mean cosmic gas density at collapse,
where $\lambda = 0.05$ (\cite{barn87}) typically.
For objects that form at $1+z > 10$,
this implies a characteristic number density of baryonic gas of at least
$300 \,{\rm cm}^{-3}$.
With this consideration in mind,
$F^{\rm cr}$ is estimated by
the fraction of matter within
the collapsed objects with mass $M$ greater than
the Jeans mass $M_{\rm J}$.
By $1+z=10$, the fraction amounts to
$50$ percent of the total mass in the universe
(\cite{fuk94}), assuming $M_{\rm J} \simeq 10^6 M_{\odot}$.
In the case that gas is considerably reheated,
$M_{\rm J}$ rises to $\simeq 10^8 M_{\odot}$ and
$F^{\rm cr}=0.3$ (\cite{fuk94}).

Gas clumps in these halos are optically thin in the relevant lines, 
and therefore
the formulation in Section \ref{subsect:formul} is valid.
For example, the one--dimensional velocity dispersion of
a $10^6 M_{\odot}$ halo at $1+z=10$ is estimated to be
$\sim 3\,{\rm km\,s^{-1}}$ using the spherical collapse model.
The size of the gas clump inside the halo with 
$n_{\rm gas} = 10^2\,{\rm cm^{-3}}$ is $\sim 10 h^{-1}{\rm pc}$.
With these numbers and 
$n_{\rm C} / n_{\rm H} = 0.01 (n_{\rm C} / n_{\rm H})_{\odot}$,
the optical depth in the CI $609\,\mu {\rm m}$ line
is estimated to be around $10^{-2}$.

Intensity fluctuations with frequency correspond to
number density fluctuations around relevant redshift:
\begin{equation}
\label{eqn:iflunflu}
   \frac{\Delta I_{\rm line}}{I_{\rm line}}
   =  \left( \frac{\Delta n}{n} \right)_{\rm source}.
\end{equation}
Even if the intensity $I_{\rm line}$ itself is smaller than the CMB
intensity $I_{\rm CMB}$, the fluctuations
$\Delta I_{\rm line} / I_{\rm tot} \cong
\Delta I_{\rm line} / I_{\rm CMB}$
can be extracted as difference between those at different frequencies.

\section{Intensity fluctuations due to linear perturbations}
\label{sect:linear}
In this section,
let us estimate the signal originated from the inhomogeneities
on a characteristic scale $\sim 1 h^{-1}\,{\rm Mpc}$
corresponding to linear perturbation.
In the next section,
we focus on smaller scales and
estimate the intensity from nonlinear clumps.
\subsection{Magnitude of the signal from carbon $609\,\mu {\rm m}$ emission}
In this and next sections, we take CI ($609 \,\mu {\rm m}$) as
an example (\cite{yam91}).
Essential discussion is common with other lines.
Since the ionization state of the universe at $z > 5$ is not known,
we begin by assuming that
most of the carbon at $1+z=10$ is in the form of CI.
The ratio of the intensity of CI ($609 \,\mu {\rm m}$) line emitted at 
$1+z=10$ to total intensity $I_{\rm tot}$ can be calculated from
equation (\ref{eqn:smoint}) with $f_u^{\rm eq} = 1/3$
(Section \ref{subsect:crnum}):
\begin{eqnarray}
\label{eqn:intenci}
   \frac{\bar{I}_{\rm line}}{I_{\rm tot}}
   & \cong &
   \frac{\bar{I}_{\rm line}}{I_{\rm CMB}} \\
   &=& 2 \times 10^{-6}
   \left( \frac{h}{0.5} \right)^{-1}
   \left( \frac{\Omega_{\rm B} h^2}{0.025} \right)
   \left( \frac{n_{\rm C}/n_{\rm H}}
   {0.01 (n_{\rm C}/n_{\rm H})_{\odot}} \right)
   \left( \frac{(n_{\rm C}/n_{\rm H})_{\odot}}{3.6 \times 10^{-4}}
   \right)
   \left( \frac{F^{\rm cr}}{0.5} \right),
\end{eqnarray}
where $h$ is the present Hubble parameter in units of
$100 \,{\rm km \,s^{-1}Mpc^{-1}}$, and $\Omega_{\rm B}$ is the present
baryon density parameter.
It is reasonable to expect that
$n_{\rm C} / n_{\rm H} = 0.01 (n_{\rm C} / n_{\rm H})_{\odot}$
by $1+z=10$.
Numerical simulation (\cite{gne96}) of nonlinear self--gravitating
clumps at high redshift shows they have a metallicity
$Z = 0.03 Z_{\odot}$ in the high density regions.
Observation of CIV absorption in Ly $\alpha$ forest clouds
also supports the above value of
$n_{\rm C} / n_{\rm H}$.
In this paper, we adopt
$(n_{{\rm C},\odot}, n_{{\rm N},\odot}, n_{{\rm O},\odot})
= (3.58, 1.12, 8.49) \times 10^{-4} n_{{\rm H}, \odot}$
for the solar abundance of carbon, nitrogen and oxygen
(\cite{and89,woo95}).

For $1+z \lesssim 15$, if most of the Carbon is in its ground states
and $n_{\rm gas} > n_{\rm gas}^{\rm cr}$,
then cooling by the CI line emission exceeds Compton cooling: the
cooling time due to the CI line emission can be estimated as
\begin{equation}
t_{\rm cool}^{\rm CI} 
  \sim \frac{(3/2)n_{\rm H}k_{\rm B} T_{\rm gas}} 
            {n_u h_{\rm P}\nu_{\rm line} A_{ul}}
  \sim 2\times10^7 \biggl(\frac{n_{\rm C}/n_{\rm H}}
                         {0.01(n_{\rm C}/n_{\rm H})_\odot}\biggr)^{\!\!-1}
                       \biggl(\frac{T_{\rm gas}}{10^3{\rm K}}\biggr)
                       \ {\rm yr},
\end{equation}
where $k_{\rm B}$ is Boltzmann's constant.
On the other hand, Compton cooling time is (\cite{lev92})
\begin{equation}
t_{\rm Compton} \sim 1\times10^8
       \biggl[ 1 + \biggl(\frac{n_e}{n_{\rm H}}\biggr)^{\!\!-1} \biggr]
                             \biggl(\frac{1+z}{10}\biggr)^{\!\!-4}\ {\rm yr},
\end{equation}
where $n_e$ is the number density of free electrons.

We can estimate number density fluctuations in the standard cold dark
matter (CDM) model with $h=0.5$, the density parameter $\Omega_0=1$,
and the present density fluctuations on $8 h^{-1}\,{\rm Mpc}$ scale
$\sigma_8=1$:
\begin{equation}
   \left( \Delta n / n \right)_{\rm atom} = \Delta M / M = 0.4
\end{equation}
on $1h^{-1} \,{\rm Mpc}$ scale at $1+z = 10$.
This has been calculated from the CDM
power spectrum $P(k)$ given by Bardeen et al. (1986).
This is a conservative estimate as the predicted fluctuation level is
even larger in vacuum--dominated or open models.
Therefore the magnitude of the signal becomes
$\Delta \bar{I}_{\rm line} / I_{\rm CMB}=1 \times 10^{-6}$ in the standard
CDM model.
Table \ref{tab:lines} summarizes predictions of intensity
fluctuations with other elements.
Note that each value has been calculated assuming that 100\% of 
the element is in the form of the neutral atom or the ion in the
first column.  Actual signal is reduced by the fraction of the
species in each element.

Since the metals are produced in dense regions,
they will be "biased" tracers of the underlying mass.
This implies that
\begin{equation}
   \left( \frac{\Delta n}{n} \right)_{\rm atom}
   = b \left( \frac{\Delta M}{M} \right),
\end{equation}
where $b$ is the bias parameter.
Gnedin and Ostriker (1996) find
a strong correlation between metallicity and density.
If $Z \propto \rho^{\alpha}$,
then
in the linear regime, this implies $b=1+\alpha$,
which enhances the signal over the values estimated 
in table \ref{tab:lines}.
\subsection{Characteristic fluctuation size}
\begin{figure}
\plotone{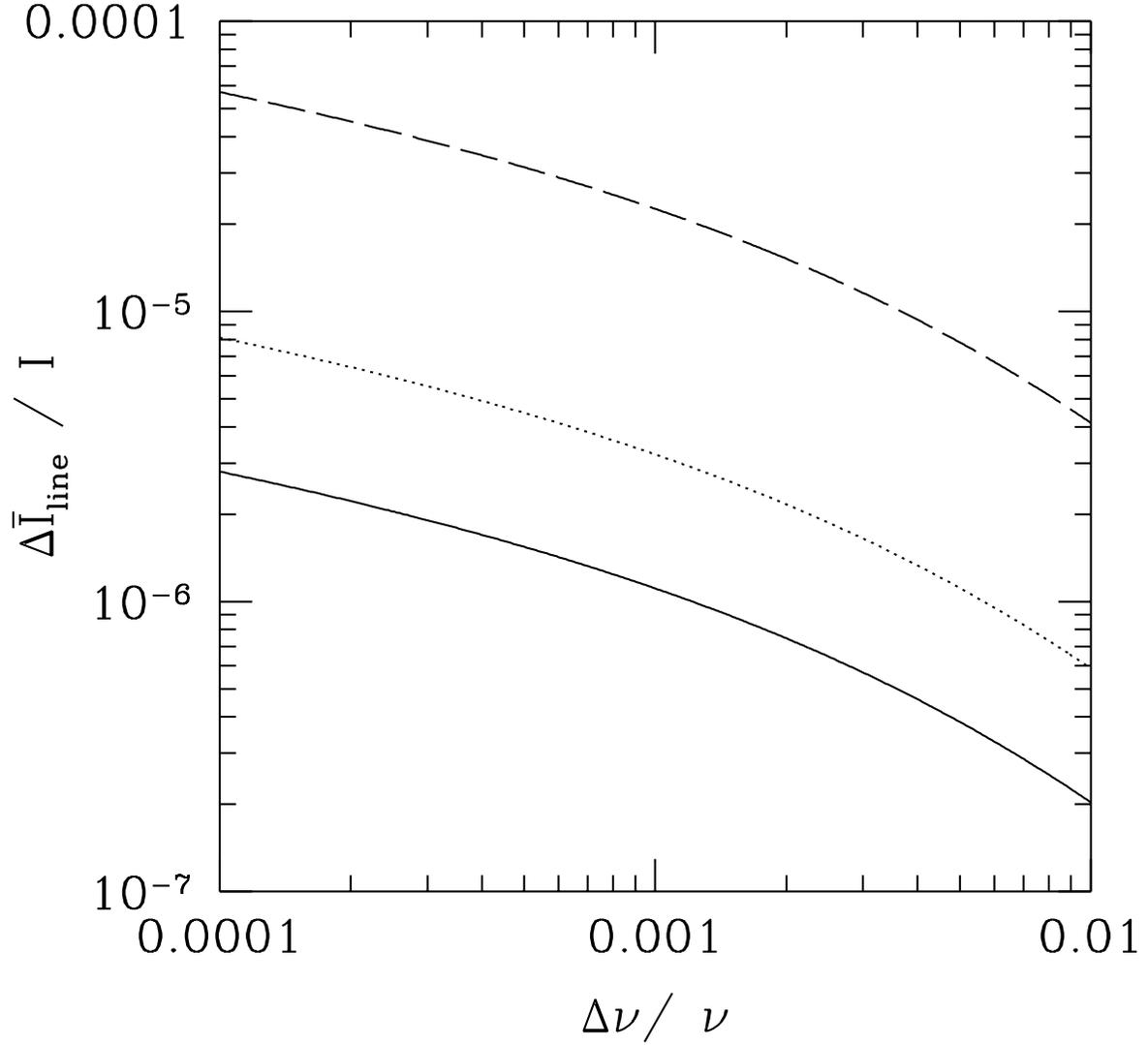}
\caption{Smoothed intensity fluctuations:
Solid and Dotted lines show the signals for
CI ($609 \,\mu {\rm m}$) and
CI ($370 \,\mu {\rm m}$) emission, respectively,
if most of the carbon is in the form of CI.
Broken line shows
the signal for
CII ($158 \,\mu {\rm m}$) emission
in the case where CII ion state is preferred.
\label{fig:f1}}
\end{figure}

Now, we investigate the resolution that we
need to detect the intensity fluctuations due to the line emission.
The comoving length scale $R_{\rm s}$ probed with frequency resolution
$\Delta \nu / \nu$ is given by
\begin{equation}
   R_{\rm s} = c\Delta \eta = \frac{c(1+z)}{H(z)} \frac{\Delta a}{a}
   = \frac{c(1+z)}{H(z)} \frac{\Delta \nu}{\nu},
\end{equation}
where $\eta$ is the conformal time, and $a=(1+z)^{-1}$.
In the Einstein--de Sitter universe,
\begin{equation}
   R_{\rm s}=cH_0^{-1}(1+z)^{-1/2}\Delta \nu/\nu
   = 0.95 \left(\frac{1+z}{10}\right)^{-1/2}
   \left(\frac{\Delta \nu/\nu}{10^{-3}}\right)
   h^{-1} \,{\rm Mpc}.
\end{equation}
An observation with a resolution $\Delta \nu/\nu$ means
that we probe the matter density inhomogeneities smoothed by
spatial thickness $R_{\rm s}$.
Figure \ref{fig:f1} shows predicted smoothed intensity fluctuations in the
standard CDM model which are
observed with given resolution $\Delta \nu / \nu$
in CI ($609 \,\mu {\rm m}$) and
CI ($370 \,\mu {\rm m}$) cases.
It also shows the signal for CII ($158 \,\mu {\rm m}$)
in the case where most carbon is in the form of CII.
Higher frequency resolution brings larger signal because of larger matter
density fluctuations.
For appropriate correspondence of frequency space to source positions,
$R_{\rm s}$ is required to satisfy the relation:
\begin{equation}
\label{eqn:ranvelsm}
   \frac{\Delta \nu}{\nu}(R_{\rm s}) >
   \frac{v_{\rm random}(R_{\rm s};z)}{c}.
\end{equation}
Here $v_{\rm random}(R_{\rm s};z)$ is the random velocity inside a
spherical region of comoving radius $R_{\rm s}$:
\begin{equation}
   v_{\rm random}(R_{\rm s};z)
   =\frac{a^2 \Omega^{0.6} H c}{2\pi^2} \int dk P(k)
   \left[ 1 - \bar{W}^2(R_{\rm s};k) \right],
\end{equation}
where $\bar{W}(R_{\rm s};k) \equiv
3(\sin{kR_{\rm s}}-kR_{\rm s}\cos{kR_{\rm s}})/(kR_{\rm s})^3$.
In the standard CDM model, the relation (\ref{eqn:ranvelsm}) yields
$\Delta \nu / \nu > 2 \times 10^{-5}$ and
$R_{\rm s} > 0.02 h^{-1}\,{\rm Mpc}$.

\placefigure{fig:f1}
Let us move on to the needed angular resolution.
To maximize observed intensity fluctuations,
we need angular resolution $\theta_{\rm res}$ corresponding to $R_{\rm s}$.
In the Einstein--de Sitter universe, this angular resolution is given by
\begin{eqnarray}
\label{eqn:angres}
  \theta_{\rm res} &=& \frac{R_{\rm s}}{\int da (a^2H)^{-1}}
  = \frac{R_{\rm s}}{2cH_0^{-1}(1-1/\sqrt{1+z})} \\
  &=& \frac{\Delta \nu/\nu}{2(\sqrt{1+z}-1)}
  = 0.79 \left( \frac{\Delta \nu/\nu}{10^{-3}}\right) \,{\rm arcmin},
\end{eqnarray}
where the last value is for $1+z=10$.
\section{Intensity fluctuations due to nonlinear clumps}
\label{sect:nonlin}
If we observe with high enough angular resolution,
we should be able to detect features
from individual clumps,
rather than
from linear matter inhomogeneities.
More concentrated objects
will be potentially stronger sources.

%
Hierarchical clustering model of structure formation in the universe
tells us how nonlinear clumps formed in the process of gravitational
instability growing.
Let us estimate typical values of the comoving size of clumps
$R_{\rm typ}$ and the mean separation among clumps
$\overline{R_{\rm typ}}$ at $1+z=10$.

The comoving number density of clumps with in the mass range
$M \sim M+dM$ at $z$ is given by the Press--Schechter (1974) formalism:
\begin{equation}
\label{eqn:psmass}
   N(M,z) dM = \sqrt{\frac{2}{\pi}} \frac{\rho_0}{M}
   \frac{\delta_{\rm c}}{D(z)}
   \left\vert \frac{1}{\sigma^2(M)}
   \frac{d\sigma (M)}{dM} \right\vert
   \exp \left[
   -\frac{\delta_{\rm c}^2}{2\sigma^2(M)D^2(z)}
   \right] dM,
\end{equation}
where $\rho_0$ is the mean comoving density of matter,
$\delta_{\rm c}$ is the overdensity threshold,
$D(z)$ is the growing function of perturbations, and
$\sigma (M)$ is the variance of density field smoothed with
mass scale $M$.
Since $\sigma (M)$ is a decreasing function of $M$,
the comoving number density (equation \ref{eqn:psmass})
exponentially falls down at large mass.
Therefore we can estimate the typical mass of clumps $M_{\rm typ}$
by a relation:
\begin{equation}
   \sigma (M_{\rm typ}) = \frac{1}{\sqrt{2}}
   \frac{\delta_{\rm c}}{D(z_{\rm re})}.
\end{equation}
Since the density of virialized objects is $18\pi^2$ times the mean
density of the universe in spherical collapse model, $R_{\rm typ}$ is
related to $M_{\rm typ}$ by
$R_{\rm typ}=(3M_{\rm typ}/4\pi \rho_0)^{1/3}(18\pi^2)^{-1/3}$.
Furthermore, the mean comoving separation
$\overline{R_{\rm typ}}=n^{-1/3}$ is given from the comoving number
density:
\begin{equation}
   n(R_{\rm typ},z) = \int_{M_{\rm typ}}^{\infty} N(M,z)dM.
\end{equation}
We make use of the fitting formula for $\sigma(M)$ in CDM models by
(\cite{kit96}):
\begin{equation}
   \sigma(M) \propto
   (1+2.208m^p-0.7668m^{2p}+0.7949m^{3p})^{-\frac{2}{9p}},
\end{equation}
which gives a good fit between $10^{-7} < m < 10^5$.
Here $p=0.0873$, and $m \equiv M (\Omega_0 h^2)^2/(10^{12} M_{\odot})$.
In the standard CDM model ($\Omega_0=1$) with
$h=0.5$ and $\sigma_8=1$,
the above equations yield
\begin{equation}
   R_{\rm typ} = 0.015 h^{-1} \,{\rm Mpc},\quad
   \overline{R_{\rm typ}} = 0.4 h^{-1} \,{\rm Mpc},
\end{equation}
at $1+z=10$.
In terms of $\Delta \nu/\nu$, these numbers correspond to
$2 \times 10^{-5}$ and $4 \times 10^{-4}$.

The one--dimensional velocity dispersion of these typical clumps
is estimated as 
\begin{equation}
  v_{\rm typ} = \sqrt{G M_{\rm typ}(1+z)/(2 R_{\rm
  typ})} = 30\,{\rm km\,s^{-1}}, 
\end{equation}
which corresponds to
$\Delta \nu/\nu = 1 \times 10^{-4}$.  
Hence the effective overdensity of the clump is
\begin{equation}
  \delta_{\rm eff} = 18\pi^2 \times
  \frac{\Delta \nu}{\nu} (R_{\rm typ}) / \frac{v_{\rm typ}}{c}
  = 30.
\end{equation}
In CI ($609 \,\mu {\rm m}$) case, for example, bumps of 
$\Delta I/I_{\rm CMB} = 1 \times 10^{-4}$ are expected to appear with
mean interval of $4 \times 10^{-4}$ in $\Delta \nu/\nu$,
if observation is made with frequency resolution 
$\Delta \nu/\nu = 1 \times 10^{-4}$ and angular resolution 
$\theta_{\rm res} \simeq 1\,{\rm arcsec}$.
Here the angular resolution is related through
equation (\ref{eqn:angres}) with
the clump size $R_{\rm typ}$.
In more conventional units in radio astronomy,
this line has intensity of
$\Delta T_{\rm line} v_{\rm typ} = 0.008 \,{\rm K\, km\, s}^{-1}$,
where $\Delta T_{\rm line}$ is the increment of brightness temperature.
In CI ($370 \,\mu {\rm m}$) case,
the amplitude of intensity fluctuations are estimated to be
$\Delta I/I_{\rm CMB} = 4 \times 10^{-4}$,
that is 
$\Delta T_{\rm line} v_{\rm typ} = 0.02 \,{\rm K\, km\, s}^{-1}$,
with the same resolution.
Alternatively,
if we assume that most carbon is in the form of CII,
the signal of CII ($158 \,\mu {\rm m}$) line is estimated to be
$\Delta I/I_{\rm CMB} = 3 \times 10^{-3}$,
that is 
$\Delta T_{\rm line} v_{\rm typ} = 0.07 \,{\rm K\, km\, s}^{-1}$.

\section{Characteristics of individual lines}
\label{sect:paremis}
The observed signal will depend upon the ionization state
and abundance of each
element.  Here, we present a simple consideration regarding the 
ionization state based on ionization potential.  More realistic
treatment of ionization of each element requires detailed computation
including temperature and density inhomogeneities of gas
and radiative transfer, which is beyond the scope of 
this paper.

CI $609\,\mu {\rm m}$ and $370\,\mu {\rm m}$ lines
have the longest wavelength among the lines discussed here.
If $1+z_s \sim 10$--$20$, they are easiest to observe from ground.
In addition, they can easily be distinguished from others.
Even if the signal is weak, one can extract the signal by
cross--correlating different portions of observed spectrum corresponding
to these two wavelengths.
The most serious concern is that
the same star that produce carbon, will quickly ionize much of it to
CII.
The ionization potential of CI, 11.3 eV, is smaller than that of
hydrogen, so we may have large mostly neutral regions containing CII.

Since carbon is so easily ionized, CII $158\,\mu {\rm m}$ line
may be a more promising candidate to be detected.
CII is the brightest submillimeter line in our galaxy (\cite{miz94}).
Unfortunately, this is a single line and its higher frequency
makes ground--based detection more challenging.

Other promising candidates are NII $204\,\mu {\rm m}$ and 
$122\,\mu {\rm m}$, and OI $146\,\mu {\rm m}$ lines.
Nitrogen is expected to be abundant only after a relatively low redshift,
since it is mainly
produced in metal--enriched stars through incomplete CNO burning.
Moreover, the ionization potential of nitrogen is 14.5 eV,
so it is more difficult to ionize than hydrogen.
On the other hand, OI abundance is supposed to peak around
the reionization epoch;
the ionization potential of oxygen is 13.6 eV, nearly
equal to that of hydrogen,
and charge exchange reaction (O$^+$ + H $\leftrightarrow$ O + H$^+$)
imply that $n_{\rm OI}/n_{\rm OII} \sim n_{\rm HI}/n_{\rm HII}$.

The other, relatively short--wavelength lines are not likely to be
observed from the ground, unless there was significant amount of metal 
production by $1+z \sim 25$--$50$.  Moreover, NIII and OIII probably
never have been dominant species in each element; the ionization
potential of NII and OII are 29.6 eV and 35.1 eV, respectively.
However, their line intensity is so strong that they may be useful
to place an interesting constraint on the state of matter
and radiation field at $1+z \sim 25$--$50$.

\section{Alternative estimate of CII intensity}
\label{sect:altcii}
%
We can make an alternative estimate of the CII $158 \mu {\rm m}$ line
intensity,
clued by the measurement that
roughly $1$ percent of the luminosity from low redshift galaxies
is emitted in the CII line (\cite{mal97}).
Though there is a possibility that the metal contamination
may have somewhat changed the fraction,
we assume that the same fraction ($1$ percent) of the luminosity
from small cloudlets at $z=z_{\rm s}$
is emitted in the CII line.

We estimate the luminosity density ${\cal L}_{\rm CII}$ of the CII
emission
instead of $h_{\rm P} \nu_{\rm line} n_u A_{ul}$
in equation (\ref{eqn:emcoeff}).
We assume that the small cloudlets evolve with a constant star
formation rate $B^{\rm com}$ per comoving unit volume
from the Salpeter type initial mass function $\phi(M)$ (\cite{sal95})
and achieve metallicity $Z(z_{\rm s})=0.01Z_{\odot}$
by $z=z_{\rm s}$.
We use the integrated abundance yields of type II supernovae tabulated by
Thielemann, Nomoto, and Hashimoto (1993)
to derive
the star formation rate
\begin{equation}
   B^{\rm com} = 0.2 \,M_{\odot}{\rm Mpc}^{-3}{\rm yr}^{-1}
   \left( \frac{1+z_{\rm s}}{10} \right)^{3/2}
   \left( \frac{h}{0.5} \right)
   \left( \frac{n_H^{\rm com}}{2.1\times 10^{-7} {\rm cm}^{-3}} \right)
   \left( \frac{\rho_{C,\odot}/\rho_{H,\odot}}{0.0043} \right)
   \left( \frac{Z/Z_{\odot}}{0.01} \right),
\end{equation}
where $n_H^{\rm com}$ is the comoving number density of hydrogen.
Then the comoving luminosity density in the B--band
is given by
\begin{equation}
   {\cal L}_B^{\rm com}
   = \int^{M_{\tau}} B^{\rm com} t_u(z_{\rm s}) \phi(M) L_B(M) dM
   + \int_{M_{\tau}} B^{\rm com} \tau(M) \phi(M) L_B(M) dM,
\end{equation}
where $t_u(z_{\rm s})$ is the age of the universe at $z=z_{\rm s}$,
$\tau(M)$ and $L_B(M)$ are the age and the B--band luminosity
of a star whose mass is $M$, respectively,
and $M_{\tau}$ is the mass which satisfies
$\tau(M_{\tau})=t_u(z_{\rm s})$.

The resultant comoving luminosity density is
${\cal L}_{\rm CII}^{\rm com} = 0.01 {\cal L}_B^{\rm com}
= 4 \times 10^7 L_{\odot} {\rm Mpc}^{-3}$ (at $1+z_{\rm s}=10$),
$7 \times 10^7 L_{\odot} {\rm Mpc}^{-3}$ (at $1+z_{\rm s}=20$).
We find that these are in good agreement with the previous estimate
$h_{\rm P} \nu_{\rm line} \bar{n}_u^{\rm com} A_{ul}
= 6 \times 10^7 L_{\odot} {\rm Mpc}^{-3}$.
We conclude that the CII intensity fluctuations estimated alternatively 
here are almost the same magnitude as the previous result
(table \ref{tab:lines}).

\section{Conclusion}
We have proposed a new and promising method to detect the first
objects: search for carbon, 
nitrogen and oxygen line emission at
$1+z=10$--$20$.
We conclude that the intensity fluctuations 
originated from the emission
are detectable in radio
band ($0.5\,{\rm mm}$--$1.2\,{\rm cm}$).
If observation is carried out with angular resolution
$\theta_{\rm res} = 1 \,{\rm arcmin}$ and spectral resolution
$\Delta \nu / \nu = 10^{-3}$,
the magnitude of the smoothed intensity fluctuations corresponding with
linear matter density fluctuations is predicted as
$\Delta \bar{I}_{\rm line} / I_{\rm tot}=
(1$ and $3) \times 10^{-6}$
(in CI $609$ and $370 \,\mu {\rm m}$ cases)
from sources at $1+z=10$, in the standard CDM model.
The predictions for other lines are
$\Delta \bar{I}_{\rm line} / I_{\rm tot} =
(20, 70, 3, 20) \times 10^{-6}$,
respectively, in
[$158$(CII),
$146$(OI),
$204$(NII),
$122$(NII)]
$\,\mu {\rm m}$
cases.

Observation with
$\theta_{\rm res} = 1 \,{\rm arcsec}$ and
$\Delta \nu / \nu = 10^{-4}$
resolve individual protogalaxies
with velocity dispersion of 
$30\,{\rm km\,s^{-1}}$.
We predict line intensities relative to the microwave background of
$\Delta I_{\rm line} / I_{\rm tot} =
(1, 4, 30, 80, 4, 30) \times 10^{-4}$
or equivalently
$\Delta T_{\rm line} v_{\rm typ} =
(0.8, 2, 7, 20, 1, 6) \times 10^{-2} \,{\rm K\, km\, s}^{-1}$
for
[$609$(CI),
$370$(CI),
$158$(CII),
$146$(OI),
$204$(NII),
$122$(NII)]
$\,\mu {\rm m}$ lines, respectively.
If detected, these features will provide useful information on matter
inhomogeneities at $1+z=10$--$20$.
They will also probe chemical evolution
and reionization time $z_{\rm re}$ itself.
\acknowledgments
 We thank John H. Black, Bruce T. Draine, Thomas Herbig,
Ryoichi Nishi, Jeremiah P. Ostriker,
Lyman A. Page, Suzanne T. Staggs, and Hajime Susa
for useful discussion.
DNS acknowledges the MAP/MIDEX
project for support.
MS and TS acknowledge support from Research
Fellowships of the Japan Society for the Promotion of Science.
\begin{table}[p]
\caption{Transition lines and intensity fluctuations:
the estimates of $\Delta \bar{I}_{\rm line} / I_{\rm tot}$
assume that the each ion is the preferred ionization state.
}
\label{tab:lines}
\begin{center}
\begin{tabular}[t]{|l|c|r|c|c|}
   \hline
   Species         & Transition                    & Wavelength      &
   $A_{ul}$        &  $\Delta \bar{I}_{\rm line} / I_{\rm tot}$ \\
                   &                               & ($\mu {\rm m}$) &
   (${\rm s}^{-1}$) & \\
   \hline
   ${\rm C}^0$     & ${}^3P_1 \to {}^3P_0$         & 609             &
   $7.93 \times 10^{-8}$ & $1 \times 10^{-6}$ \\
   ${\rm C}^0$     & ${}^3P_2 \to {}^3P_1$         & 370             &
   $2.68 \times 10^{-7}$ & $3 \times 10^{-6}$ \\
   ${\rm C}^+$     & ${}^2P_{3/2} \to {}^2P_{1/2}$ & 158             &
   $2.36 \times 10^{-6}$ & $2 \times 10^{-5}$ \\
   ${\rm N}^+$     & ${}^3P_1 \to {}^3P_0$         & 204             &
   $2.13 \times 10^{-6}$ & $3 \times 10^{-6}$ \\
   ${\rm N}^+$     & ${}^3P_2 \to {}^3P_1$         & 122             &
   $7.48 \times 10^{-6}$ & $2 \times 10^{-5}$ \\
   ${\rm N}^{++}$  & ${}^2P_{3/2} \to {}^2P_{1/2}$ &  57.3           &
   $4.77 \times 10^{-5}$ & $2 \times 10^{-3}$ \\
   ${\rm O}^0$     & ${}^3P_1 \to {}^3P_2$         &  63.2           &
   $8.95 \times 10^{-5}$ & $9 \times 10^{-3}$ \\
   ${\rm O}^0$     & ${}^3P_0 \to {}^3P_1$         & 146             &
   $1.70 \times 10^{-5}$ & $7 \times 10^{-5}$ \\
   ${\rm O}^{++}$  & ${}^3P_2 \to {}^3P_1$         &  51.8           &
   $9.75 \times 10^{-5}$ & $5 \times 10^{-2}$ \\
   ${\rm O}^{++}$  & ${}^3P_1 \to {}^3P_0$         &  88.4           &
   $2.62 \times 10^{-5}$ & $7 \times 10^{-4}$ \\
   \hline
\end{tabular}
\end{center}
\end{table}

\end{document}